\documentclass[conference]{IEEEtran}

\makeatletter
\def\endthebibliography{%
	\def\@noitemerr{\@latex@warning{Empty `thebibliography' environment}}%
	\endlist
}
\makeatother

\usepackage[utf8]{inputenc}
\usepackage[english]{babel}

\usepackage{amsthm}
\usepackage{xpatch}
\usepackage{color}
\makeatletter
\xpatchcmd{\proof}{\@addpunct{.}}{\@addpunct{:}}{}{}
\makeatother
\usepackage{cite}
\usepackage{amssymb}
\usepackage{amsmath}
\usepackage{graphicx}
\usepackage{stfloats}
\usepackage{algorithm}
\usepackage{algorithmic}

\usepackage[figurename=Fig.]{caption}
\usepackage{comment}

\usepackage{enumerate}
\DeclareMathOperator*{\argmax}{arg\,max}
\DeclareMathOperator*{\argmin}{arg\,min}
\renewcommand\IEEEkeywordsname{Index Terms}

\IEEEoverridecommandlockouts
\begin{document}
%
\title{Distributed Computation of A Posteriori Bit Likelihood Ratios in Cell-Free Massive MIMO}


%
\author{Zakir Hussain Shaik$^{*}$, Emil Bj{\"o}rnson$^{*\dagger}$ and Erik G. Larsson$^{*}$\\
	$^{*}$Department of Electrical Engineering (ISY), Link{\"o}ping University, Link{\"o}ping, Sweden\\$^{\dagger}$Department of Computer Science, KTH Royal Institute of Technology, Stockholm, Sweden\\
	Email: \{zakir.hussain.shaik, erik.g.larsson\}@liu.se, emilbjo@kth.se \thanks{This work was partially supported by the Swedish Research Council (VR) and ELLIIT.}}


\maketitle
\begin{abstract}
This paper presents a novel strategy to decentralize the soft detection procedure in an uplink cell-free massive multiple-input-multiple-output network. We propose efficient approaches to compute the a posteriori probability-per-bit, exactly or approximately, when having a sequential fronthaul. More precisely, each access point (AP) in the network computes partial sufficient statistics locally, fuses it with received partial statistics from another AP, and then forward the result to the next AP. Once the sufficient statistics reach the central processing unit, it performs the soft demodulation by computing the log-likelihood ratio (LLR) per bit, and then a channel decoding algorithm (e.g., a Turbo decoder) is utilized to decode the bits. We derive the distributed computation of LLR analytically.
\end{abstract}

%
\IEEEpeerreviewmaketitle

\vspace{-1.2mm}
\section{Introduction}
Cell-free massive multiple-input-multiple-output (mMIMO) is envisaged to be one of the beyond 5G technologies~\cite{ZhangBeyond5G}. It is a decentralized implementation of mMIMO with no cell boundaries as opposed to the traditional cellular networks~\cite{Hien,interdonato2019ubiquitous,cfMIMObookEmil,ZhangCellFree}. In cell-free mMIMO, many access points (APs) are deployed in a geographical area to serve the user equipments (UEs) jointly whereby providing macro-diversity gain~\cite{interdonato2019ubiquitous}. An AP is a circuitry that comprises antenna elements and the signal processing units required to operate them locally. Different from other distributed MIMO technologies, the operating regime has many more APs than UEs, but each AP has much fewer antennas than there are UEs and, thus, must cooperate with other APs to manage interference. The topology of the interconnections between the APs is arbitrary, e.g., star, daisy-chain, etc., depending on the application.

The original idea of a cell-free network was to have a star topology, i.e., each AP has a dedicated fronthaul (a link between two nodes) to the CPU \cite{Hien}. In this network, all the APs estimate the channel locally and make a local estimate of the data. Then all the APs share the estimated data with the CPU, which decodes the information signals. In \cite{Nayebi2016a,mMIMObookEmil}, different implementation architectures with varying levels of cooperation between the APs and CPU are studied. In the centralized implementation, the CPU has global information and thereby always has the superior performance, say in terms of spectral efficiency (SE), over distributed implementations with partial information at the CPU. On the other hand, this type of implementation increases the overall fronthaul capacity (amount of information shared to the CPU) and also the cost of deployment if a wired implementation is considered. One possible solution to address these issues is by decentralizing the network operating using efficient algorithms that can distribute the signal processing computation and ensure minimal loss in the performance, such as SE and bit-error-rate (BER), compared to the centralized network implementation. A few other benefits of distributed signal processing are system reliability, scalability of the network to setups with many APs and UEs, and privacy. Some possible choices for decentralized topologies are sequential, tree network, etc., where the APs process its information locally and forward partial information to the CPU \cite{shaik2020mmseoptimal,Bertilsson}. 
For example, \cite{shaik2020mmseoptimal} studied the sequential topology for a so-called radio stripes network. In a radio stripe network \cite{interdonato2019ubiquitous}, the APs are sequentially connected (i.e., using a daisy-chain architecture) and share the same cables for front-haul and power supply.

The algorithms developed for decentralizing mMIMO in the literature can be adopted in a cell-free mMIMO network, but these algorithms do not take advantage of cooperation among APs effectively. In the literature, the works focusing on the decentralized implementation of mMIMO are:  \cite{decntZF} where the authors designed a decentralized implementation of an approximate zero-forcing (ZF) precoding; \cite{decentZF2} that explored various algorithms to decentralize ZF precoding in uplink and downlink with different algorithms providing a trade-off between fronthaul signaling and latency; on the similar lines decentralized ZF methods are also studied in the context of large intelligent surfaces, and one such example is \cite{LIS1}. A few other relevant works on the decentralized implementation of mMIMO are discussed \cite{JeonDecent,LiDecent,ShiraziniaDecent,Bertilsson,Atzeni}. A recent work that focused on cell-free mMIMO networks with distributed algorithms is \cite{shaik2020mmseoptimal}, in which the authors developed a sequentially distributed algorithm in a radio stripes network that achieves the maximum SE.

\textit{Contribution}: \textcolor{black}{In practice, system design should ensure that the performance at the bit level is ensured over SE or other soft estimate metrics like the mean-square error (MSE) because the information is transmitted in bits with finite length codewords.} We focus on establishing an approach to decode the information at the bit level by computing the likelihood of a bit. There is no prior work that computes the posterior bit likelihood in a distributed network. We investigate the computation of the likelihood of the transmitted bits and analytically derive a method  to compute the log-likelihood ratio (LLR) of the bits in decentralized networks, specifically in sequentially connected or tree networks. The new method requires less fronthaul signaling than a centralized implementation. Besides computing distributed LLRs, the important features of the proposed algorithm are that it holds for imperfections in the channel state information (CSI) and is scalable with respect to the number of APs in the network. This work essentially shows that the optimal non-linear detector (in the sense of bit-error-rate) can be decentralized.

\textit{Notations:} 
The superscripts $(\cdot)^*,~(\cdot)^T,$ and $(\cdot)^H$ denote conjugate, transpose, and Hermitian transpose, respectively. The $N\times N$ identity matrix is  $\mathbf{I}_N$. A block diagonal matrix is denoted by $\mathrm{bldiag}(\mathbf{A}_1,\ldots,\mathbf{A}_N)$ with square matrices $\mathbf{A}_1,\ldots,\mathbf{A}_N$. 
We denote expectation by $\mathbb{E}\{\cdot\}$. We use $\mathbf{z} \sim \mathcal{CN}\left(\mathbf{0},\mathbf{C}\right)$ to denote a multi-variate circularly symmetric complex Gaussian random vector with zero mean and covariance matrix $\mathbf{C}$. We denote the probability density function (PDF) of a 
random variable $x$ by $f(x)$. 

\begin{figure}[!t]
	\centering
	\includegraphics[width=.7\linewidth]{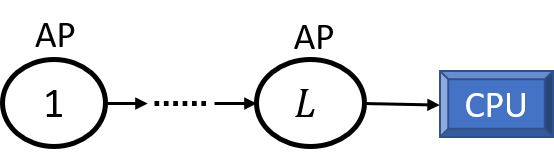}
	\caption{Sequential architecture of cell-free mMIMO network.}
	\label{fig:TreeStr}
\end{figure}
\section{System Model And Channel Estimation}\label{SystemModel}
We consider a cell-free mMIMO network comprising $L$ APs connected in a daisy-chain architecture, each equipped with $N\geq 1$ antennas. Without loss of generality, the fronthaul connection is assumed to have the sequence AP $1$ - AP $2$ - AP $3$ - $\cdots$ - AP $L$ - CPU, where the CPU is located at the end of the network as shown in Fig.~\ref{fig:TreeStr}. A radio stripes network \cite{interdonato2019ubiquitous} is one example of such an architecture. The algorithm proposed in this paper can also be extended to a tree network \cite{Bertilsson}.

There are $K\ll NL$ single-antenna UEs distributed arbitrarily in the considered coverage area. We use the block fading channel model with the coherence block length of $\tau_c$ channel uses. The channel between AP $l$ and UE $k$ is denoted by $\mathbf{h}_{kl} \in \mathbb{C}^{N}$. In each block, an independent realization is drawn from a correlated Rayleigh fading distribution as
\begin{equation}
\mathbf{h}_{kl} \sim \mathcal{CN} \left(\mathbf{0},\mathbf{R}_{kl}\right),
\end{equation}
where $\mathbf{R}_{kl} \in \mathbb{C}^{N \times N}$ is the spatial correlation matrix, which attributes the channel spatial correlation characteristics and large-scale fading. \textcolor{black}{We assume APs are sufficient distant apart to assume that there is no correlation between APs}. We also assume that the spatial correlation matrices $\{\mathbf{R}_{kl}\}$ are known at all the APs.

This paper analyzes an uplink scenario consisting of $\tau_p$ and $\tau_c - \tau_p$ channel uses for the pilot transmission to estimate the channels and the payload data, respectively.  
\subsection{Channel Estimation}
We assume that there are $\tau_p$ mutually orthogonal $\tau_p$-length pilot vector signals $\boldsymbol{\phi}_1,~ \boldsymbol{\phi}_2,~ \ldots,~ \boldsymbol{\phi}_{\tau_p}$ with $\Vert \boldsymbol{\phi}_{k} \Vert^2=\tau_p$, which are used for channel estimation. When $K>\tau_p$, more than one UE is assigned with the same pilot, which causes pilot contamination. We let the pilot assigned to UE $k$, where $ k=1,\ldots, K$, to be indexed as $t_k \in \{1,\ldots,\tau_p\}$ and the set $\mathcal{S}_k = \{ i : t_i = t_k\}$ accounts for those UEs assigned with the same pilot as that of UE $k$.
The received signal at AP $l$ during the pilot transmission is $\mathbf{Y}_l^p \in \mathbb{C}^{N \times \tau_p}$, given by
\begin{equation}
\mathbf{Y}_l^p = \sum_{i=1}^{K} \sqrt{p_i}\mathbf{h}_{il}\boldsymbol{\phi}_{t_i}^T + \mathbf{N}_l, \vspace{-2mm}
\end{equation}
where $p_i \geq 0$ is the transmit power of UE $i$, $\mathbf{N}_l \in \mathbb{C}^{N \times \tau_p}$ is the additive white Gaussian receiver  modeled with independent entries distributed as $\mathcal{CN}(0,\sigma^2)$ with $\sigma^2$ being the noise power. The minimum mean square error (MMSE) estimate $\widehat{\mathbf{h}}_{kl}\in \mathbb{C}^{N\times1}$ of the channel is given by \cite{mMIMObookEmil}
\begin{equation}
\begin{aligned}
\widehat{\mathbf{h}}_{kl}=\sqrt{p_k \tau_p}\mathbf{R}_{kl}\boldsymbol{\Psi}_{t_k l}^{-1}\mathbf{y}_{t_k l}^p,
\end{aligned}
\end{equation}\vspace{-4mm}
where
{\small
\begin{align}
\mathbf{y}_{t_k l}^p & = \mathbf{Y}_l^p\frac{\boldsymbol{\phi}_{t_k}^{*}}{\sqrt{\tau_p}} 
=  \sum_{i \in \mathcal{S}_k}\sqrt{p_i\tau_p}\mathbf{h}_{il} + \mathbf{n}_{t_k l}, \\
\boldsymbol{\Psi}_{t_k l} & = \sum_{i \in \mathcal{S}_k} \tau_p p_i\mathbf{R}_{il} + \sigma^2 \mathbf{I}_N
\end{align}
}
are the despreaded signal and its covariance matrix, respectively. Here, $\mathbf{n}_{t_k l} \triangleq \mathbf{N}_l {\boldsymbol{\phi}_{t_k}^*}/{\sqrt{\tau_p}} \sim \mathcal{CN}\left(\mathbf{0},\sigma^2\mathbf{I}_{N}\right)$ is the effective noise. An important consequence of MMSE estimation is that the estimate $\widehat{\mathbf{h}}_{kl} \sim \mathcal{CN}(\mathbf{0},\widehat{\mathbf{R}}_{kl})$ and the estimation error $\widetilde{\mathbf{h}}_{kl} = \mathbf{h}_{kl} -\widehat{\mathbf{h}}_{kl} \sim \mathcal{CN}(\mathbf{0},\widetilde{\mathbf{R}}_{kl})$ are independent, with $\widehat{\mathbf{R}}_{kl} =p_k\tau_p\mathbf{R}_{kl}\boldsymbol{\Psi}_{t_k l}^{-1}\mathbf{R}_{kl},\ \widetilde{\mathbf{R}}_{kl} = \mathbf{R}_{kl} -\widehat{\mathbf{R}}_{kl}$ as the respective covariance matrices.
\subsection{Uplink Payload Transmission}
During the uplink payload transmission phase, UE $k$ transmits data symbols  $s_k \in \mathcal{M}$ from the signal constellation alphabet $\mathcal{M}=\{a_1,\ldots,a_M\}$ comprising $M$ symbols.  We assume the symbols transmitted by UE $k$ are chosen independently of UE $m$ for $k \neq m$. The received signal $\mathbf{y}_l \in \mathbb{C}^N$ at AP $l$ is 
\begin{equation}\label{ULrecSig2}
\mathbf{y}_l = \mathbf{H}_{l}\mathbf{s} + \mathbf{n}_l, \vspace{-2mm}
\end{equation}
where $\mathbf{H}_l = [\mathbf{h}_{1l},\mathbf{h}_{2l},\ldots,\mathbf{h}_{Kl}]\in \mathbb{C}^{N\times K}$ is the channel matrix, $\mathbf{s}= [s_{1},s_{2},\ldots,s_{K}]^T\in \mathcal{M}^K$ is the transmit signal vector, and $\mathbf{n}_l \sim \mathcal{CN}\left(\mathbf{0},\sigma^2 \mathbf{I}_N\right)$ is the AP $l$ receiver noise. We assume that symbols transmitted by UEs are equally likely, i.e., $\mathbf{s}$ is uniformly distributed over $ \mathcal{M}^K$.

Let $\mathbf{H}_{l} = \widehat{\mathbf{H}}_l + \widetilde{\mathbf{H}}_l$ with $ \widehat{\mathbf{H}}_l = [\widehat{\mathbf{h}}_{1l},\widehat{\mathbf{h}}_{2l},\ldots,\widehat{\mathbf{h}}_{Kl}]$ being the channel matrix estimate and $ \widetilde{\mathbf{H}}_l = [\widetilde{\mathbf{h}}_{1l},\widetilde{\mathbf{h}}_{2l},\ldots,\widetilde{\mathbf{h}}_{Kl}]$ is the channel estimation error matrix with $\widetilde{\mathbf{h}}_{kl} = \mathbf{h}_{kl}-\widehat{\mathbf{h}}_{kl}$. Accordingly, \eqref{ULrecSig2} is equivalent to
\begin{equation}
\begin{aligned}
\mathbf{y}_l = \widehat{\mathbf{H}}_l\mathbf{s} + \mathbf{w}_l,
\end{aligned}
\end{equation}
where  $\mathbf{w}_l = \widetilde{\mathbf{H}}_l\mathbf{s} + \mathbf{n}_l$ can be thought of as a colored noise term at AP $l$. An important attribute of $\mathbf{w}_l$, which we will exploit later, is that for given $\mathbf{s}$ it is conditionally Gaussian with zero conditional mean and conditional covariance, given by
\begin{equation}\label{covWs}
\mathbf{\Sigma}_{l|\mathbf{s}} = \mathbb{E}\{\mathbf{w}_l \mathbf{w}_l^H | \mathbf{s}\} = \sum_{i=1}^{K}\vert s_i\vert^2\widetilde{\mathbf{R}}_{il} + \sigma^2\mathbf{I}_N.
\end{equation}
Besides being conditionally Gaussian, $\mathbf{w}_l$ is also conditionally independent to $\mathbf{w}_m,\ l\neq m$ for a given $\mathbf{s}$. 

\section{Decentralized Detection}
The important task of the receiver is to detect the most probable transmitted symbol sequences based on the information available at the receiver. Designing reliable MIMO detectors poses a huge challenge due to the complexity involved in the implementation. We refer to \cite{LarssonDetectionTut,Albreem} for detailed reviews of MIMO detection methods. In the literature, there are broadly speaking two types of detectors for the detection of transmitted symbols (or bits): hard-decision and soft-decision detectors. Examples of hard-decision detectors include maximum likelihood (ML) and maximum a posteriori (MAP) methods. On the other hand, the soft-decision detectors quantify how reliable are the decisions on the symbols (or bits) in the information-carrying signals. In most cases, the soft-decision detectors have superior performance over the hard-decision detectors \cite{proakisBook}. 

The standard MIMO detection methods are appropriate for systems with co-located antennas, where the receiver can operate close to the antenna array and, thus, have access to all the CSI that exist in the system. However, the \textcolor{black}{standard non-distributed methods} are not suitable for cell-free mMIMO where the CSI is distributed between many APs, each estimating a subset of the channels, observing a subset of the received data signals, and having local processing capabilities. In principle, all the APs could send their information to the CPU, which can implement a standard detection method, but this requires a lot of fronthaul signaling and is not making use of the local processors.
We want to take advantage of the distributed computation capabilities to develop distributed MIMO detection algorithms that also require less fronthaul signaling. We start by briefly discussing the implementation of the MAP hard-decision detector in a distributed network and then consider soft-detectors (specifically, computation of bit-likelihood ratios), which is the main focus of this paper. 
 
We first describe a centralized detector that will serve as our benchmark. A centralized cell-free network with $L$ APs operates in two phases. In the first phase, all the APs send the pilot signals to the CPU from which it estimates the channel, and then the CPU receives the data signal from which it forms the following augmented received signal
\begin{equation}\label{centReceiveSig}
\mathbf{z}_L = \widehat{\mathbf{G}}_L\mathbf{s}+\overline{\mathbf{w}}_L\, 
\end{equation}
with 
{\small$\mathbf{z}_L = \left[\mathbf{y}_1^H,\ldots,\mathbf{y}_L^H\right]^H,\ 
\widehat{\mathbf{G}}_L =  \left[\widehat{\mathbf{H}}_1^H,\ldots,\widehat{\mathbf{H}}_L^H\right]^H,\ 
\overline{\mathbf{w}}_L\ = \left[\mathbf{w}_1^H,\ldots,\mathbf{w}_L^H\right]^H,$ } where $\mathbf{z}_L \in \mathbb{C}^{NL\times 1}$ is the received signal for all APs, $\widehat{\mathbf{G}}_L \in \mathbb{C}^{NL \times K}$ is the matrix with channel estimates, and $\overline{\mathbf{w}}_L \in \mathbb{C}^{NL\times 1}$ is the colored noise. The noise vector $\overline{\mathbf{w}}_L$ is conditionally Gaussian for a given $\mathbf{s}$ with zero mean and has the conditional covariance  $
\mathbf{K}_{L|\mathbf{s}}= \textrm{bldiag}\left(\mathbf{\Sigma}_{1|\mathbf{s}},\ldots,\mathbf{\Sigma}_{L|\mathbf{s}}\right)$.

\subsection{MAP Detector for Hard Detection}
The MAP detector for a centralized cell-free network is defined as follows:
{\small
	\begin{align}
	\widehat{\mathbf{s}}_L &= \argmax_{\mathbf{s} \in \mathcal{M}^{K}} f\left(\mathbf{s}|\mathbf{z}_L, \widehat{\mathbf{G}}_L\right)\label{mapRule}\\
	& \overset{(a)}{=} \argmin_{\mathbf{s} \in \mathcal{M}^{K}} \left\Vert \mathbf{K}_{L|\mathbf{s}}^{-1/2}\left(\mathbf{z}_L-\widehat{\mathbf{G}}_L\mathbf{s}\right)\right\Vert^2 + \ln\left(\textrm{det} \left(\mathbf{K}_{L|\mathbf{s}}\right)\right)\notag,
	\end{align}
}
where $(a)$ is obtained by applying Bayes' rule along with utilizing the conditional Gaussian distribution of $\mathbf{z}_L$ and uniform distribution of $\mathbf{s}$, and then taking the logarithm of the argument and simplifying. 
Interestingly, the last expression in \eqref{mapRule} can be computed in a sequential manner:
{\small
\begin{equation}\label{mapRule2}
\begin{aligned}
\widehat{\mathbf{s}}_L &= \argmin_{\mathbf{s} \in \mathcal{M}^{K}} \sum_{l=1}^{L} \left[\left\Vert \mathbf{\Sigma}_{l|\mathbf{s}}^{-1/2}\left(\mathbf{y}_l-\widehat{\mathbf{H}}_l\mathbf{s}\right) \right\Vert^2 + \ln(\textrm{det} (\mathbf{\Sigma}_{l|\mathbf{s}}))\right]\\
&= \argmin_{\mathbf{s} \in \mathcal{M}^{K}} \left[b_{L|\mathbf{s}} + \mathbf{s}^H\mathbf{M}_{L|\mathbf{s}} \mathbf{s} - 2 \mathcal{R}\left\{\mathbf{a}_{L|\mathbf{s}}^H\mathbf{s}\right\}+c_{L|\mathbf{s}}\right],
\end{aligned}
\end{equation}
} 
where the variables appearing on the second row can be computed  iteratively as follows:
{
\begin{equation} \label{varsAP}
\begin{aligned}
b_{l|\mathbf{s}} & = b_{(l-1)|\mathbf{s}}+ \Vert \mathbf{r}_{l|\mathbf{s}}\Vert^2,\\
\mathbf{M}_{l|\mathbf{s}} &= \mathbf{M}_{(l-1)|\mathbf{s}} + \widehat{\mathbf{C}}_{l|\mathbf{s}}^H\widehat{\mathbf{C}}_{l|\mathbf{s}},\\
\mathbf{a}_{l|\mathbf{s}} & = \mathbf{a}_{(l-1)|\mathbf{s}}+\widehat{\mathbf{C}}_{l|\mathbf{s}}^H\mathbf{r}_{l|\mathbf{s}},\\
c_{l|\mathbf{s}} &= c_{(l-1)|\mathbf{s}} + \ln(\textrm{det} (\mathbf{\Sigma}_{l|\mathbf{s}})),
\end{aligned}
\end{equation}
}where $\mathbf{r}_{l|\mathbf{s}} = \mathbf{\Sigma}_{l|\mathbf{s}}^{-1/2}\mathbf{y}_l,\ 
\widehat{\mathbf{C}}_{l|\mathbf{s}} = \mathbf{\Sigma}_{l|\mathbf{s}}^{-1/2}\widehat{\mathbf{H}}_l$ for $l=1,\dots,L$. The computation is initiated by $\mathbf{M}_{0|\mathbf{s}}$ being a $K \times  K$ matrix with zeros,  $b_{0|\mathbf{s}}=0,\ c_{0|\mathbf{s}}=0$, and $\mathbf{a}_{0|\mathbf{s}}$ is a $K \times 1$ zero vector. 
Hence, the exact MAP detector can be implemented in a sequential manner that fits the sequential fronthaul architecture shown in Fig.~\ref{fig:TreeStr}. AP $l$ computes the variables $\{b_{l|\mathbf{s}},\mathbf{M}_{l|\mathbf{s}},\mathbf{a}_{l|\mathbf{s}},c_{l|\mathbf{s}}\}$
according to \eqref{varsAP} and forwards them to AP $(l+1)$. When the CPU receives $\{b_{L|\mathbf{s}},\mathbf{M}_{L|\mathbf{s}},\mathbf{a}_{L|\mathbf{s}},c_{L|\mathbf{s}}\}$ from the last AP, it can compute the cost function in \eqref{mapRule2} and make the MAP detection. 

The proposed sequential implementation limits the information that must flow from the APs towards the CPU. 
However, a main issue is that $\{b_{l|\mathbf{s}},\mathbf{M}_{l|\mathbf{s}},\mathbf{a}_{l|\mathbf{s}},c_{l|\mathbf{s}}\}$ depend on $\mathbf{s}$ and, thus, must be computed  for all possible combinations of $\mathbf{s} \in \mathcal{M}^K$ making its practical implementation difficult. The dependence enters into the expression through the conditional covariance $\mathbf{\Sigma}_{l|\mathbf{s}}$, defined in \eqref{covWs}. If phase-shift keying (PSK) is utilized so that $\vert s_i\vert^2=p_i$ for all $s_i \in \mathcal{M}$, then the dependence on $\mathbf{s}$ disappears since
\begin{equation}\label{covWs_approx}
\mathbf{\Sigma}_{l|\mathbf{s}} = \sum_{i=1}^{K} p_i\widetilde{\mathbf{R}}_{il} + \sigma^2\mathbf{I}_N.\vspace{-2mm}
\end{equation}
We can also employ this as an approximation for modulations with amplitude variations.

Using \eqref{covWs_approx}, we now introduce a set of variables that do not depend on $\mathbf{s}$: $\mathbf{\Sigma}_{l} = \mathbf{\Sigma}_{l|\mathbf{s}}$, 
$\mathbf{r}_{l} =  \mathbf{r}_{l|\mathbf{s}}$, 
$\mathbf{a}_{l} =  \mathbf{a}_{l|\mathbf{s}}$,
$\widehat{\mathbf{C}}_{l} = \widehat{\mathbf{C}}_{l|\mathbf{s}}$, $\mathbf{M}_{l} =\mathbf{M}_{l|\mathbf{s}}$, 
for all $l = \{1,2,\ldots, L\}$ under the condition that $\vert s_i\vert^2=p_i, \forall i\in\{1,2,\ldots, K\}$. By removing the terms that do not explicitly depend on $\mathbf{s}$ in \eqref{mapRule2}, the MAP detection at the CPU can now be computed as
\begin{equation}\label{rsMAP_approx}
\widehat{\mathbf{s}}_L = \argmin_{\mathbf{s} \in \mathcal{M}^{K}}  \left[\mathbf{s}^H\mathbf{M}_L\mathbf{s} - 2 \mathcal{R}\left\{\mathbf{a}_L^H\mathbf{s}\right\}\right].
\end{equation}
\vspace{-10mm}
\subsection{Fronthaul signaling comparison} We will now quantify the difference in fronthaul signaling between a centralized implementation and the proposed sequential implementation, based on the simplification in \eqref{covWs_approx}.
We measure the fronthaul signaling in terms of the number of real symbols shared in the link connecting AP $L$ with the CPU. In a centralized implementation, to compute \eqref{mapRule}, each AP has to send the following information to the CPU: $(i)$ $\mathbf{y}_l$ which amounts to $2N$ real symbols in every channel use and $(ii)$ pilot signals $2N\tau_p$ per coherence block. This sums up to 
$2NL\tau_c$ real symbols per coherence block and all the fronthaul traffic must pass through AP $L$.
With the proposed sequential implementation, AP $L$ has to forward: $(i)$ $\mathbf{a}_L$ for every channel use, amounting to $2K$ real symbols and $(ii)$ $\mathbf{M}_L$ once in every coherence block, containing $K^2$ real symbols. This sums up to $2K(\tau_c-\tau_p) + K^2$ real symbols per coherence block. Note that the fronthaul signaling of the centralized implementation grows linearly with $L$, the number of APs. On the other hand, the fronthaul requirement in sequential topology textcolor{black}{ more efficiently distributed in the links and} with the proposed algorithm, \textcolor{black}{the fronthaul in the link connecting the CPU and AP $L$} is independent of the number of APs, making it scalable for use in networks with many APs.

\begin{figure}[!t]
	\centering
	\includegraphics[height=4.5cm, width=8cm]{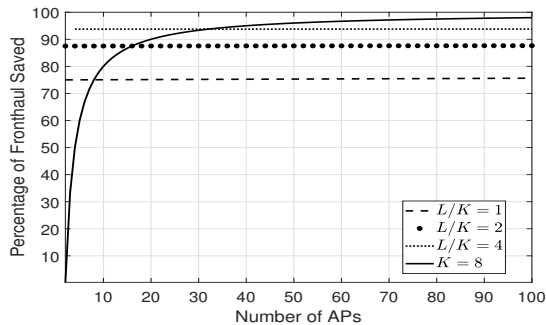}
	\caption{Percentage of {fronthaul saved} by the proposed algorithm compared to centralized implementation.}
	\label{figFronthaul}
\end{figure}

Fig.~\ref{figFronthaul} shows the percentage of fronthaul signaling that is saved by the sequential implementation over a centralized implementation for $\tau_c = 2000,\ \tau_p = K,\ N = 4$. We observe that the fronthaul saving is large and almost constant as we increase $L$ for a fixed $L/K$ ratio. If $K$ is fixed, the fronthaul signaling saved increases rapidly with $L$, e.g., for $L=24,\ N =4,\ K =8,\ \tau_c =2000,\ \tau_p =8$, the sequential implementation requires approximately $91$\% less fronthaul signaling than the centralized implementation. 

\subsection{LLR Calculation for Soft Detection}
The transmitted vector $\mathbf{s}$ contains bits that represent some underlying information. In practice, a long sequence of bits corresponds to a codeword from a channel code, thus we do not want to make a hard detection of the individual bits but of the entire codeword. To this end, the receiver should compute the likelihood of the bits and provide it as soft input to the decoding algorithm of the channel code (e.g., a turbo decoder). We will develop a sequential algorithm for that case.

Let the number of bits required to represent each symbol in $\mathbf{s}$ be $m=\log_2(M)$ (e.g., $m=2$ represents Quadrature phase-shift keying (QPSK)), therefore the vector $\mathbf{s}$ has a total of $mK$ bits. We also assume that these bits are independent (can be achieved in practice with interleaver) and equally likely. We denote these bits as $b_1,\ldots,b_{mK}$. The associated a priori LLR of each bit $b_i$ is given by
\begin{equation}
\mathcal{L}(b_i)=\ln\left(\frac{P(b_i=1)}{P(b_i=0)}\right).
\end{equation}
The posterior LLR for a centralized implementation using the conditional density function in \eqref{mapRule} $(b)$ and with the
assumption that $\mathbf{s}$ is uniformly distributed is given by:
\begin{equation}\label{centLLR}
\mathcal{L}(b_i|\mathbf{z}_L) = \ln\left(\frac{\sum_{\mathbf{s}:b_i(\mathbf{s})=1}f(\mathbf{z}_L|\mathbf{s},\widehat{\mathbf{G}}_L)}{\sum_{\mathbf{s}:b_i(\mathbf{s})=0}f(\mathbf{z}_L|\mathbf{s},\widehat{\mathbf{G}}_L)}\right).
\end{equation}
The notation $\mathbf{s}:b_i(\mathbf{s})=\alpha$ means the set of all vectors $\mathbf{s}$ for which the $i$th bit is $\alpha$ i.e., $b_i(\mathbf{s})=\alpha$. After the likelihood values of the bits are computed using \eqref{centLLR}, the channel decoder decodes the data bits. However, it is known that the computational complexity of \eqref{centLLR} increases exponentially with the increase in the number of UEs \cite{vcirkic2014efficient}, this is because the summation in \eqref{centLLR} contains $2^{mK}$ terms. To address this problem, many sub-optimal solutions exist and one such method is called max-log approximation \cite{llrErik}. In this method, each of the sums in \eqref{centLLR} is approximated with their largest term. Accordingly, \eqref{centLLR} is written as
{\small
\begin{equation}\label{centLLRMaxLog}
\begin{aligned}
\mathcal{L}(b_i|\mathbf{z}_L) \overset{(a)}{=} & \ln\left(\frac{\max_{\mathbf{s}:b_i(\mathbf{s})=1}f(\mathbf{z}_L|\mathbf{s},\widehat{\mathbf{G}}_L)}{\max_{\mathbf{s}:b_i(\mathbf{s})=0}f(\mathbf{z}_L|\mathbf{s},\widehat{\mathbf{G}}_L)}\right)\\
=& \min_{\mathbf{s}:b_i(\mathbf{s})=0}\left\Vert\mathbf{K}_{L|\mathbf{s}}^{-1/2}\left(\mathbf{z}_L-\widehat{\mathbf{G}}_L\mathbf{s}\right)\right\Vert^2-\\
&\min_{\mathbf{s}:b_i(\mathbf{s})=1}\left\Vert\mathbf{K}_{L|\mathbf{s}}^{-1/2}\left(\mathbf{z}_L-\widehat{\mathbf{G}}_L\mathbf{s}\right)\right\Vert^2,
\end{aligned}
\end{equation}
}
where $(a)$ follows from \eqref{covWs_approx} and, thus, is exact for PSK modulations and an approximation otherwise. We will now show how to implement both exact and log-max approximate LLR computation in a sequential manner that fits a cell-free mMIMO network of the kind in Fig.~\ref{fig:TreeStr}. We define the following notation, to simplify the LLR analytical expressions:
\begin{equation}
\begin{aligned} 
\psi^{'} \left(\mathbf{s},\mathbf{M}_L,\mathbf{a}_{L|\mathbf{s}}\right) &= \textrm{exp}\left(-\mathbf{s}^H\mathbf{M}_L\mathbf{s}+2\mathcal{R}\left\{\mathbf{a}_{L|\mathbf{s}}\mathbf{s}\right\}\right),\\
\psi \left(\mathbf{s},\mathbf{M}_L,\mathbf{a}_{L}\right) &= \textrm{exp}\left(-\mathbf{s}^H\mathbf{M}_L\mathbf{s}+2\mathcal{R}\left\{\mathbf{a}_{L}\mathbf{s}\right\}\right).
\end{aligned}
\end{equation}
The exact posterior LLR computation in \eqref{centLLR} can be implemented in an distributed manner by re-writing \eqref{centLLR} as
{\small
\begin{equation}\label{rsLLR}
	\mathcal{L}(b_i|\mathbf{z}_L)= \ln\left(\frac{\sum_{\mathbf{s}:b_i(\mathbf{s})=1}d_{L|\mathbf{s}}\textrm{exp}\left(-b_{L|\mathbf{s}}\right)\psi^{'}(\mathbf{s},\mathbf{M}_L,\mathbf{a}_{L|\mathbf{s}})}{\sum_{\mathbf{s}:b_i(\mathbf{s})=0}d_{L|\mathbf{s}}\textrm{exp}\left(-b_{L|\mathbf{s}}\right)\psi^{'}(\mathbf{s},\mathbf{M}_L,\mathbf{a}_{L|\mathbf{s}})}\right),
\end{equation}
}
where
\begin{align}
d_{l|\mathbf{s}} &= d_{(l-1)|\mathbf{s}} \textrm{det}(\mathbf{\Sigma}_{l|\mathbf{s}})^{-1},\ \ d_{0|\mathbf{s}} = 1;\ l = \{1,\ldots,L\}.
\end{align}
The LLR computation in \eqref{centLLR} is equivalent to that  in \eqref{rsLLR}. Hence, we have obtained a sequential way to compute the a posteriori bit LLRs in a cell-free mMIMO system. For the CPU to compute the exact LLR, each AP has to compute and forward the terms given in \eqref{varsAP}.
The main bottleneck in \eqref{rsLLR} is the dependency of the conditional covariance on the transmitted symbols in every channel use, having the same computational complexity as discussed for the MAP rule. This can be simplified by making use of the property in \eqref{covWs_approx}, which is exact for PSK modulation and otherwise an approximation. Thus, the LLR computation can be simplified as
{\small
 \begin{equation}\label{llr_maxlog_approx}
 \begin{aligned}
\mathcal{L}(b_i|\mathbf{z}_L)= \ln\left(\frac{\sum_{\mathbf{s}:b_i(\mathbf{s})=1}\psi(\mathbf{s},\mathbf{M}_L,\mathbf{a}_{L})}{\sum_{\mathbf{s}:b_i(\mathbf{s})=0}\psi(\mathbf{s},\mathbf{M}_L,\mathbf{a}_{L})}\right).
 \end{aligned}
 \end{equation}
}

{\small
\begin{algorithm}[t]
	\caption{Decentralized MAP/Soft-detectors given in \eqref{rsMAP_approx}, \eqref{llr_maxlog_approx} and \eqref{maxLog_rs_approx} for sequential network.} 
	\begin{algorithmic} \label{Algo1}
		\STATE  1. \textbf{Initialize}: $\mathbf{M}_0=\mathbf{0},\ \mathbf{a}_0 = \mathbf{0}$;
		\STATE 2. \textbf{for} $l = 1:L$
		\STATE \quad\ (i)  Compute $\mathbf{M}_{l} = \mathbf{M}_{(l-1)} + \widehat{\mathbf{C}}_{l}^H\widehat{\mathbf{C}}_{l}$ 
		\STATE \quad\ (ii) Compute $\mathbf{a}_{l} = \mathbf{a}_{(l-1)} + \widehat{\mathbf{C}}_{l}^H\mathbf{r}_{l}$
		\STATE \quad \textbf{end}
		\STATE 3. \textbf{Output}: Compute the MAP detector/soft-detectors expressions given in \eqref{rsMAP_approx}, \eqref{llr_maxlog_approx} and \eqref{maxLog_rs_approx}.
	\end{algorithmic}
\end{algorithm}
}
Similarly, the max-log approximation can be computed in a decentralized manner as follows
{\small
\begin{equation}\label{maxLog_rs}
\begin{aligned}
\mathcal{L}(b_i|\mathbf{z}_L)=\ln\left(\frac{\max_{\mathbf{s}:b_i(\mathbf{s})=1}d_{L|\mathbf{s}}\textrm{exp}\left(-b_{L|\mathbf{s}}\right)\psi^{'}(\mathbf{s},\mathbf{M}_L,\mathbf{a}_{L|\mathbf{s}})}{\max_{\mathbf{s}:b_i(\mathbf{s})=0}d_{L|\mathbf{s}}\textrm{exp}\left(-b_{L|\mathbf{s}}\right)\psi^{'}(\mathbf{s},\mathbf{M}_L,\mathbf{a}_{L|\mathbf{s}})}\right).
\end{aligned}
\end{equation}
}
\hspace{-.4em}Similar to \eqref{llr_maxlog_approx}, the complexity involved in max-log computation for a decentralized network can be reduced by making the assumption in \eqref{covWs_approx}, thus \eqref{maxLog_rs} becomes
{\small
\begin{equation}\label{maxLog_rs_approx}
\begin{aligned}
\mathcal{L}(b_i|\mathbf{z}_L)=&\min_{\mathbf{s}:b_i(\mathbf{s})=1}\ln(\psi(\mathbf{s},\mathbf{M}_L,\mathbf{a}_{L}))\\&-\min_{\mathbf{s}:b_i(\mathbf{s})=0}\ln(\psi(\mathbf{s},\mathbf{M}_L,\mathbf{a}_{L})).
\end{aligned}
\end{equation}
}
\vspace{-1mm}
A pseudo-code for implementing the proposed sequential hard and soft detectors is given in Algorithm \ref{Algo1}.

To summarize, the computation of the exact and max-log approximation, given in \eqref{centLLR} and \eqref{centLLRMaxLog}, respectively, can be implemented in a decentralized manner as given in \eqref{rsLLR} and \eqref{llr_maxlog_approx}, respectively. This implementation fits a cell-free mMIMO network with a sequential fronthaul.
Moreover, a relaxed version with lower computational complexity considering the assumption in \eqref{covWs_approx} for both distributed exact and max-log approximated of LLR are given in \eqref{llr_maxlog_approx} and \eqref{maxLog_rs_approx} respectively. One of the drawbacks of the proposed method is latency which grows linearly with $L$. Nevertheless, the advantages of radio stripes implementation outweigh the drawbacks in non-latency critical applications. 

\vspace{-1mm}
\section{Conclusion}
This paper introduces a novel method to compute a posteriori bit LLRs analytically in a decentralized manner in cell-free mMIMO networks when there is a sequential fronthaul, as in radio stripes networks. The proposed method has two important practical features. First, it is designed for imperfect CSI scenarios. Second, the fronthaul load required is independent of number of APs, i.e., the algorithm is scalable with respect to the number of APs. While previous works focused on the distributed computation of MMSE-based algorithms, this paper focuses on the distributed computation of bit likelihood which is an important quantity of interest practically.

\vspace{-1.1mm}
\bibliographystyle{IEEEtran}
\bibliography{reff}

\end{document}